\documentclass[%
showpacs,preprintnumbers,
 amsmath,amssymb,
 aps,
prb,
]{revtex4}

\usepackage{graphicx}
\usepackage{dcolumn}
\usepackage{bm}

\usepackage{amsmath}    
\usepackage{graphicx}   
\usepackage{verbatim}   
\usepackage{color}      
\usepackage{subfigure}  
\usepackage{hyperref}   
\usepackage{mathrsfs}

\begin{document}


\title{A class of scalable parallel and vectorizable pseudorandom number generators based on non-cryptographic RSA exponentiation ciphers}

\author{Jetanat Datephanyawat}
\email{jetanat.oat@gmail.com}
\author{Paul D. Beale}
\email{paul.beale@colorado.edu}
\affiliation{University of Colorado Boulder}

\date{\today}

\begin{abstract}
Parallel supercomputer-based Monte Carlo and stochastic simulations require pseudorandom number generators that can produce distinct pseudorandom streams across many independent processes. 
We propose a scalable class of parallel and vectorizable pseudorandom number generators  based on a non-cryptographic version of the RSA public-key exponentiation cipher. 
Our method generates uniformly distributed IEEE double-precision floating point pseudorandom sequences on $[0,1)$ by encrypting pseudorandom sequences of 64-bit integer messages by modular exponentiation.  
The advantages of the method are:  the method is parallelizable by parameterization with each pseudorandom number generator instance derived from an independent 64-bit composite modulus, the method is fully scalable on massively parallel computing clusters because of the millions of available 32-bit prime numbers, the seeding and initialization of the independent streams is simple, the periods of the independent instances are all different and greater than  $8.5\times 10^{37}$, 
and the method passes a battery of intrastream and interstream correlation tests. The calculations in each instance can be vectorized using steam splitting and can produce more than $10^8$ pseudorandom numbers per second on each multicore CPU.
\end{abstract}

\maketitle


\section{\label{sec:Introduction}Introduction}

Parallel supercomputer-based Monte Carlo and stochastic simulations require pseudorandom number generators that can produce distinct pseudorandom streams across many independent processes. 
We have developed a class of scalable parallel and vectorizable pseudorandom number generators 
for use in massively parallel supercomputer applications. The method we propose is
based on a non-cryptographic version of the Rivest, Shamir and Adleman (RSA) public key exponentiation cipher.\cite{rsa1978, FergusonSchneierKohno2010,Schneier1994, Koshy2002, Silverman2006, Koblitz1987} 
The method creates pseudorandom streams by encrypting sequences of 64-bit integer plaintext \textit{messages} $m_k$ into \textit{ciphertexts} $c_k$ using the transformation 
\begin{align}\label{eqn:encrypt2}
c_k&= m_k^e \thinspace\textrm{mod}\thinspace  n .
\end{align}
Each generator instance is based on an independent composite modulus $n=p q$, where $p$ and $q$ are 32-bit primes, and the exponent $e$ is a small odd number.  
Here and throughout, $x=y \thinspace\textrm{mod}\thinspace z$ means $x$ is the remainder of $y$ upon division by $z$,  with $0 \leq x < z$.
Note that this is \textit{not} a cryptographically secure pseudorandom generator, which would need to operate on integers that are thousands of bits long. 
The algorithm is fully scalable by parametrization on parallel supercomputers since each node can be assigned independent pairs of primes. The algorithm is vectorizable, and can generate more than $10^8$ pseudorandom numbers per second on each multicore supercomputer node.

The pseudorandom number generator algorithm described here cycles through a sequence of integer messages $m_k$ with $k=0,1,2,\ldots$ uniformly selected from $\mathbb{Z}_n=[0\thinspace . \thinspace . \thinspace n-1]$. The encryption step in equation \eqref{eqn:encrypt2} then gives a 
sequence of pseudorandom ciphertexts $c_k$ that is uniformly distributed on $\mathbb{Z}_n$. Uniformly distributed double-precision floating point pseudorandom numbers $r_k$ on the real interval $[0 \thinspace ,\thinspace 1)$ are formed with a floating point division: $r_k= c_k / n$. 
Since $(a b c)  \thinspace\textrm{mod}\thinspace  n = \left( a\thinspace (bc \thinspace\textrm{mod}\thinspace  n)\right)\thinspace\textrm{mod}\thinspace  n$, repeated squaring and multiplying can be used to evaluate the exponentiation \eqref{eqn:encrypt2} in less than $2 \log_2 e$ modular multiplications.

Most pseudorandom number generators generate the next pseudorandom integer from either the previous pseudorandom integer in the sequence, or by operating on two or more pseudorandom integers from earlier in the sequence. In our method, the pseudorandom sequence arises from the encryption of a  sequence of integer messages. In this way, it is similar to cryptographically secure pseudorandom number generators,\cite{Schneier1994, FergusonSchneierKohno2010} and pseudorandom number generators based on block ciphers.\cite{NumericalRecipes1992,Rijmen,Random123,U01Test} The quality of the pseudorandom sequence produced by our method is based on modular exponentiation being a good one-way cryptographic function. \cite{rsa1978, Schneier1994, FergusonSchneierKohno2010,Shparlinski2000} 

Large-scale parallel programs that use pseudorandom numbers should utilize an algorithm that is scalably parallel. Otherwise, different processes risk sampling overlapping pseudorandom subsequences which would give results that are not statistically independent. 
Coddington\cite{Coddington} recommends parallel pseudorandom number generators should have the following characteristics (paraphrased here):
\begin{itemize}

\item The method should produce uncorrelated pseudorandom numbers in any number of processes, and pass a battery of stringent tests of randomness within each process and between processes.

\item The algorithm should have a period that is provably far longer than a single process can exhaust in any conceivable run.

\item The method should be able to create one instance, or a scalably large number of independent instances.

\item  To assist in debugging, the user should be able to seed the algorithm to give independent reproducible results in any number of processes.

\item The code should be portable across wide range of computer architectures.

\item The generator should have fast initialization and execution speeds that use limited memory, and each process should run independently once initialized.

\end{itemize}

Two qualitatively different schemes have been used to create scalable systems of pseudorandom number generators: stream splitting and parameterization.\cite{BaukeMertens2007} Parallelization by stream splitting is based on a single pseudorandom number generator with an extremely long period, with parallelization accomplished by subdividing the full period into non-overlapping subsequences. 
By contrast, parallelization by parameterization produces independent pseudorandom sequences by using generators chosen from a set of independent generators, and assigning a different fixed set of parameters to the generator in each process. 
The most widely used classes of parallel pseudorandom number generators are based on the lagged Fibonacci method,\cite{mascagni1995a, mascagni1995b, MascagniSrinivasan2000, sprng, MersenneTwister,knuth}  
which calculates the next pseudorandom integer from two previous integers in the sequence  
$s_k = (s_{k-t} \circledcirc s_{k-r}) \thinspace\textrm{mod}\thinspace 2^m$, where $\circledcirc$ is one of the operations bitwise exclusive or, addition, subtraction, or multiplication, and  $t<r$ are integer parameters chosen based on primitive polynomials modulo 2 that implement a Galois finite  field with order $2^r$.\cite{knuth,ZierlerBrillhart} The bitwise exclusive-or algorithms with $m=1$ have periods of $P=2^r -1$,  additive, subtractive, and word-wise exclusive-or algorithms have periods of $P=(2^r-1)2^{m-1}$, and multiplicative algorithms have periods of $P=(2^r-1)2^{m-2}$. The parameter $r$ is typically chosen in the range of several hundred to several thousand. The state of the generator is defined by a table of $r$ $m$-bit integers, which represent the most recent pseudorandom integers in the sequence. Parallel implementation of these algorithms can be accomplished by either stream splitting or parameterization.\cite{mascagni1995a, mascagni1995b, MascagniSrinivasan2000, sprng} 

Our method is parallelized by parameterization by assigning a unique modulus $n=p q$ to each process. 
The number of independent streams is limited only by the number of prime pairs in the range defined by the implementation. Furthermore, the generator on each independent process can be 
vectorized by taking advantage of  stream splitting to calculate a vector of pseudorandom numbers, which greatly speeds the calculation on vector processors or multicore CPUs.

\section{\label{sec:RSA}RSA public key encryption}

Asymmetric or public-key cryptography was first publicly proposed Ralph Merkle.\cite{Merkle1978} 
Whitfield Diffie and Martin Hellman\cite{DiffieHellman1976} were the first to publish a practical algorithm for key exchange based on modular exponentiation in a prime field, and Ron Rivest, Adi Shamir and Leonard Adleman (RSA)\cite{rsa1978} published their public key cryptosystem paper in 1978. (It is important to note that all of these methods were discovered earlier by British GCHQ mathematicians in highly classified work: James Ellis proposed the idea of asymmetric ciphers, or what he called \textit{non-secret encryption}, Malcolm Williamson developed a key exchange method identical to Diffie--Hellman, and Clifford Cocks developed a version of the RSA algorithm. These discoveries were not revealed publicly until their work was declassified in 1997.\cite{SinghCodeBook})

Our generator mimics the manner that RSA is used in practice to establish a secure communications channel between Alice and Bob.\cite{FergusonSchneierKohno2010}  
Alice wants to send an encrypted message to Bob even though they have never met to securely exchange a secret symmetric encryption/decryption key. 
Bob first creates a public key consisting of a composite number $n=p q$ where $p$ and $q$ are two large secret prime numbers. Bob publicly shares the product $n$ and a small exponent $e$ that is coprime to $p-1$ and $q-1$. Alice uses Bob's public key to encrypt her message $m$ into ciphertext $c=m^e\thinspace\textrm{mod}\thinspace n$, and sends the ciphertext to Bob over an open channel. Bob can decrypt the ciphertext using $m=c^d \thinspace\textrm{mod}\thinspace n$, where Bob's decryption exponent $d$ can be determined from $e$ and the two secret primes.
RSA is usually used to securely share a secret symmetric key $K$. Alice encrypts a random key $K$ using Bob's public key, and sends the ciphertext to Bob who  decrypts the key. Both Alice and Bob can then use their shared secret key $K$ in a fast symmetric encryption algorithm.

Alice and Bob assume that an eavesdropper Eve will be able to intercept the ciphertext $c$. The security of RSA is based on both multiplication and modular exponentiation being good one-way functions, i.e. multiplying and modular exponentiation are \textit{easy}, while factoring and solving the discrete logarithm problem are \textit{hard}.\cite{FergusonSchneierKohno2010,Schneier1994,Koshy2002,Silverman2006} 
Easy and hard are distinguished by whether or not a function can be calculated 
in \textit{polynomial time}, i.e. in a time proportional to some power of $\log n$.
Cryptographic security currently requires that $n$ should be thousands of bits long. 
No known classical algorithm can factor large composites or calculate discrete logarithms in a polynomial time. It is these problems that a multi-thousand qubit quantum computer could potentially crack.\cite{Shor} 



\section{\label{sec:Mathematics}Number Theory}


For every prime number $p$, the set of integers $\mathbb{Z}_p = [0 \thinspace . \thinspace . \thinspace p-1]$ forms a finite field, i.e. $\mathbb{Z}_p$ is closed under addition and subtraction modulo $p$, and the set of nonzero elements $\mathbb{Z}_p^*= [1 \thinspace . \thinspace . \thinspace p-1]$ forms a group that is closed under multiplication and division modulo $p$.  Division is defined since for every integer $a\in\mathbb{Z}_p^*$ there exists a unique multiplicative inverse $a^{-1}\in\mathbb{Z}_p^*$ such that $a a^{-1}\thinspace\textrm{mod}\thinspace  p = 1$ . 

In the RSA method a composite $n=p q$ is created from two large primes $p$ and $q$, and 
any message $m \in \mathbb{Z}_n$ can be encrypted into a unique ciphertext $c\in \mathbb{Z}_n$ by exponentiation mod $n$:
\begin{align}\label{eqn:encrypt}
c&= m^e \thinspace\textrm{mod}\thinspace  n .
\end{align}
The ciphertext can be decrypted using a decryption exponent $d$:\cite{rsa1978, Schneier1994, Koshy2002, Silverman2006, Koblitz1987}
\begin{align}\label{eqn:decrypt}
m&= c^d \thinspace\textrm{mod}\thinspace  n .
\end{align}
The decryption exponent $d$ exists and is unique if $e$ and $(p-1)(q-1)$ are co-prime. 
Decryption is based on Fermat's little theorem:\cite{Koshy2002, Silverman2006, Koblitz1987} 
for any prime $p$ and for all $m \in \mathbb{Z}_p^*$, $m^{p-1}\thinspace\textrm{mod}\thinspace p= 1$. 
For the case of composite moduli of the form $n=p q$, 
the generalization of Fermat's little theorem is for all $m$ co-prime to $n$, $m^{\phi(n)} \thinspace \textrm{mod} \thinspace n = 1$, where $\phi(n)=(p-1)(q-1)$ is Euler's totient function, the number of elements in $\mathbb{Z}_n^*$ that are coprime to $n$.
The decryption exponent is given by 
\begin{align}\label{eqn:RSAd}
d=e^{-1} \thinspace \textrm{mod} \thinspace \phi(n) ,
\end{align}
i.e. $de = 1 + u \phi(n)$ for some integer $u$. 
The decryption exponent $d$  can be easily calculated using the extended Euclidean algorithm 
by anyone who knows $e$, $p$ and $q$.\cite{Schneier1994, Koshy2002, Silverman2006, Koblitz1987} 
The validity of equation \eqref{eqn:decrypt} is demonstrated as follows:
\begin{align}\label{eqn:decryptproof}
c^d \thinspace \textrm{mod} \thinspace n = m^{d e} \thinspace \textrm{mod} \thinspace n =
m^{1+u {\phi(n)}} \thinspace \textrm{mod} \thinspace n =
(m (m^{\phi(n)})^u) \thinspace \textrm{mod} \thinspace n =
m \thinspace \textrm{mod} \thinspace n = m .
\end{align}
Encryption and decryption are 
one-to-one mappings of $\mathbb{Z}_n$ onto $\mathbb{Z}_n$, so any message sequence 
$\{m_k\}$ that uniformly samples $\mathbb{Z}_n$ will produce a ciphertext sequence $\{c_k\}$ that uniformly samples $\mathbb{Z}_n$. 

The Chinese remainder theorem (CRT)\cite{FergusonSchneierKohno2010,Schneier1994,Koshy2002,Silverman2006,Koblitz1987}
can be used to speed up the modular exponentiations modulo $n$.
 Every integer $m \in \mathbb{Z}_n$ with $n=p q$ can be uniquely represented in terms of two numbers $m_p\in \mathbb{Z}_{p}$ and $m_q\in \mathbb{Z}_{q}$ given by
\begin{subequations}
\begin{align}
&m_p=m \thinspace\textrm{mod}\thinspace p ,\\ 
&m_q=m \thinspace\textrm{mod}\thinspace q,
\end{align}
\end{subequations}
and the value of $m$ can be recovered from $m_p$ and $m_q$ using Garner's formula:
\begin{align}
m =  (((m_p - m_q) (q^{-1}\thinspace\textrm{mod}\thinspace p)) \thinspace\textrm{mod}\thinspace p ) q + m_q.
\end{align}
The exponentiation $c=m^e\thinspace\textrm{mod}\thinspace n$ can then be accomplished 
by exponentiating $m_p$ and $m_q$ and using Garner's formula: 
\begin{subequations}\label{eqn:CRTimplementation}
\begin{align}
c_p& = m_p^e \thinspace\textrm{mod}\thinspace p ,\\
c_q& = m_q^e \thinspace\textrm{mod}\thinspace q ,\\
c\thinspace &=  (((c_p - c_q) (q^{-1}\thinspace\textrm{mod}\thinspace p)) \thinspace\textrm{mod}\thinspace p ) q +c_q.
\end{align}
\end{subequations}
Since we choose $p$ and $q$ to be 32-bit primes, the multiplications and exponentiations in equations \eqref{eqn:CRTimplementation} can be accomplished using fast native 64-bit arithmetic. (In RSA cryptographic applications, the CRT-based speedup of the exponentiations can only be used in the decryption step, since only Bob knows $p$ and $q$.)

\section{\label{sec:PseudorandomSkips}Pseudorandom skips}

The sequence of messages to be encrypted can be expressed in terms of an integer skip sequence  $\{s_k\}$ 
chosen from $\mathbb{Z}_n$:
\begin{align}
m_k&= (m_{k-1}+s_k) \thinspace\textrm{mod}\thinspace n,
\end{align}
Note that if the skip sequence uniformly and \textit{randomly} (not pseudorandomly) samples $\mathbb{Z}_n$, then the sequence of messages $m_k$ forms a uniform random sequence on $\mathbb{Z}_n$. Each message is, in effect, a one-time pad encryption of the previous message.\cite{Schneier1994}  
We will approximate this by choosing the skips $s_k$ pseudorandomly. 

In encryption it is essential to avoid \textit{cribs}, i.e. messages that result in easily decoded ciphertexts. For example, the messages $m=0,1,n-1$ are cribs for all allowed exponents $e$ since $m^e \thinspace\textrm{mod}\thinspace n = 0,1,n-1$, respectively. 
RSA-based cryptographic applications often use encryption exponents as small as $e=3$ or $5$ for efficiency.\cite{FergusonSchneierKohno2010} Messages with $m^e < n$ and $(n-m)^e < n$ result in trivially decodable ciphertexts, so exponents $e < \log_2 n$ result in additional cribs. 
In cryptographic applications, messages are randomly padded\cite{Schneier1994, FergusonSchneierKohno2010} 
to avoid cribs. 
For our purposes, it is not necessary to eliminate cribs, since they would appear in any long random sequence of messages, but rather to prevent correlated sequences of cribs.
Our goal is to select a simple skip pattern that ensures a uniform sampling of the set of all messages, avoids correlated cribs, is computationally fast, has a long period, and allows the use of small encryption exponents. 

The simplest skip sequence that uniformly samples $\mathbb{Z}_n$ is the unit skip, i.e. $s_k = 1$ for all $k$. Encryption of this sequence gives a block cipher operating in counter mode 
which is used in some cryptographically secure pseudorandom number generators.\cite{Schneier1994,FergusonSchneierKohno2010} 
The message sequence is $m_k=(m_0 + k) \thinspace\textrm{mod}\thinspace n$, with period $P=n$. 
Except for the cribs near $m=0$ and $n$, pseudorandom sequences derived from a unit skip  
empirically pass the U01 battery of statistical correlations tests\cite{U01Test} for exponents $e \geq 9$.
A constant skip $s_k = b$ with $1<b<n-1$, can eliminate sequential cribs. 
but
the pattern produced by constant skips should not be substantially better than 
the unit skip pattern since 
\begin{align}\label{constantskipsymmetry}
(kb)^e\thinspace\textrm{mod}\thinspace n=\left( (b^e \thinspace\textrm{mod}\thinspace n)(k^e \thinspace\textrm{mod}\thinspace n)\right) \thinspace\textrm{mod}\thinspace n,
\end{align}
is just a constant multiplier permutation of the unit skip sequence. 

We choose a skip sequence produced by a prime number linear congruential pseudorandom number 
generator:\cite{knuth,lecuyer1988,LecuyerBlouinCouture1993,SezginSezgin2013}
\begin{align}\label{congruentialgenerator1}
s_k= a s_{k-1} \thinspace\textrm{mod}\thinspace Q= s_0 a^k \thinspace\textrm{mod}\thinspace Q ,
\end{align}
with prime modulus $Q$. The multiplier $a$ is chosen to be a primitive root $\textrm{mod}\thinspace Q$\cite{Koshy2002, Silverman2006, Koblitz1987} that  
delivers a full period, well-tested
pseudorandom sequence.

If $n$ is coprime to $Q$ and $Q-1$, then the period of the message sequence is $P=(Q-1)n$. 
This is shown by noting that 
since $a$ is  
a primitive root mod $Q$, after $Q-1$ steps $s_k$ will have cycled through every value in $\mathbb{Z}_Q^*$, so $s_{k+Q-1}=s_k$ and $m_{k+Q-1}=\left( m_k + \sum_{s=1}^{Q-1}{s} \right) \thinspace\textrm{mod}\thinspace n = ( m_k + Q(Q-1)/2 ) \thinspace\textrm{mod}\thinspace n$.  
If $\textrm{gcd}(Q(Q-1)/2,n)=1$, then $b=Q(Q-1)/2 \thinspace\textrm{mod}\thinspace n \neq 0$.
The skip and message values shifted forward by $(Q-1)u$ steps are given by 
$s_{k+(Q-1)u}=s_k$ and $m_{k+(Q-1)u}=(m_k+ub)\thinspace\textrm{mod}\thinspace n$. Therefore, every subsequence of messages of length $Q-1$ is different from every other subsequence. 
Since  $s_{k+(Q-1)n}=s_k$ and $m_{k+(Q-1)n}=m_k$, the period of the generator is $P=(Q-1)n$. 

Using Fermat's little theorem, the state of the generator after $k=u(Q-1) + v$ steps, with $u=\lfloor k/(Q-1)\rfloor$ and $v=k\thinspace\textrm{mod}\thinspace (Q-1)$, is given by 
\begin{subequations}\label{pseudoskip}
\begin{align}
s_k &= s_0 a^k \thinspace\textrm{mod}\thinspace Q = s_0 a^v \thinspace\textrm{mod}\thinspace Q 
= a^{v_0+v} \thinspace\textrm{mod}\thinspace Q ,\\
m_k &= 
\left( m_0 + \sum_{j=1}^k{ s_0 a^j \thinspace\textrm{mod}\thinspace Q } \right)\thinspace\textrm{mod}\thinspace n , \notag \\
&= \left( m_0 +ub + \sum_{j=1}^v{ a^{v_0+j} \thinspace\textrm{mod}\thinspace Q } \right)\thinspace\textrm{mod}\thinspace n , 
\label{pseudoskipb}  \\
c_k &= m_k^e \thinspace\textrm{mod}\thinspace n , 
\end{align}
\end{subequations}
where $s_0=a^{v_0} \thinspace\textrm{mod}\thinspace Q$. 
Even though the message and ciphertext sequence can be expressed in closed form, calculating the values of $m_k$ and $c_k$ for large $k$ 
or determining $k$ from $(m,s)$, 
is computationally laborious 
unless $k$ is close to a multiple of $Q-1$. 

The method has the following properties:

\begin{itemize}


\item The algorithm is based on elementary cryptography and number theory, and satisfies all of Coddington's criteria.\cite{Coddington}

\item Using a pseudorandom skip extends the period of the generator to $P=(Q-1)n$, and provides a uniform sampling of ciphertexts over the full period of the generator. Each message $m \in \mathbb{Z}_n$, and hence each ciphertext $c \in \mathbb{Z}_n$, will appear exactly $Q-1$ times in the full-period sequence. 








\item The method is parallelizable by both parameterization and stream splitting, with each independent process derived from a distinct composite modulus  $n=p q$. The method is fully scalable on massively parallel supercomputers due to the millions of 32-bit primes. 

\item Pseudorandom sequences that result from different moduli are independent, and have different  periods.



\item The seeding and initialization of the independent streams is simple. 

\item The state of each generator is defined by a few fixed integer parameters $\{n=p q,e,Q,a\}$, and a few integer state values $\{m,s,c\}$ that are updated during each call to the generator.





\item By using 32-bit primitive roots mod $Q$ for the skip generator, 32-bit primes $p$ and $q$, the CRT, and small exponents $e$, the implementation below requires only a few fast native 64-bit integer operations per pseudorandom number.




\item The method passes a battery of strong randomness tests, within each stream and between streams.

\end{itemize}

\section{\label{sec:Implementation}Implementation in 64 bits }





We choose the pseudorandom skip modulus to be $Q=2^{63}-25$, the largest prime less than $2^{63}$. This choice will deliver a maximum period $P=(Q-1)n$ since $Q-1$ is coprime to all primes in $[2^{31},2^{32}]$.
Equation \eqref{congruentialgenerator1} can be implemented using only 64-bit arithmetic if the primitive root is chosen from the restricted set of values 32-bit values.\cite{WichmanHill1982,BratleyFoxSchrage1987,lecuyer1988} This can be shown by expressing $Q$ 
in the form 
$Q=a q_1 + q_2$, where $q_1=\lfloor Q / a \rfloor$ and $q_2=Q \thinspace\textrm{mod}\thinspace a$:
\begin{align}\label{restrictedprimitiveroots}
a s \thinspace\textrm{mod}\thinspace Q &= (a s - \lfloor s/q_1\rfloor Q)\thinspace\textrm{mod}\thinspace Q \notag \\
&=  (a s - \lfloor s/q_1\rfloor (a q_1 + q_2))\thinspace\textrm{mod}\thinspace Q \notag \\
&=(a (s - \lfloor s/q_1\rfloor q_1) - q_2 \lfloor s/q_1\rfloor))\thinspace\textrm{mod}\thinspace Q \notag \\
&= (a (s \thinspace\textrm{mod}\thinspace q_1) - q_2 \lfloor s/q_1\rfloor ))\thinspace\textrm{mod}\thinspace Q .
\end{align}
By restricting the choice of primitive roots $a$ to the range $[2^{31},2^{32}]$ that also give $q_2<q_1$, then 
the intermediate results 
$s_1 = a (s \thinspace\textrm{mod}\thinspace q_1)$ and $s_2 = q_2 \lfloor s/q_1\rfloor$ in the final line of equation \eqref{restrictedprimitiveroots} are both less than $Q$ which allows a simple 64-bit   implementation. L'Ecuyer, Blouin, and Couture,\cite{LecuyerBlouinCouture1993} and  
Sezgin and Sezgin\cite{SezginSezgin2013} give a handful of restricted primitive roots 
that have good spectral test properties and pass all U01 Crush and BigCrush tests.\cite{U01Test}
By using restricted primitive roots, 32-bit primes, and the CRT, the entire calculation can be implemented using fast native unsigned 64-bit integer arithmetic. 
The algorithm uses three pre-calculated values $q_1=\lfloor Q / a \rfloor$, $q_2=Q \thinspace\textrm{mod}\thinspace a$, and $q^{-1}\thinspace\textrm{mod}\thinspace p$.
The pseudocode for generating the next double-precision floating point pseudorandom number $r$ on the interval $[0\thinspace ,\thinspace 1)$ is given by:
\begin{subequations}\label{algorithm1}
\begin{align}
&s_1 := a \thinspace (s \thinspace\textrm{mod}\thinspace q_1) ,\label{eqn:WH1}\\
&s_2 := q_2 \thinspace \lfloor s/q_1\rfloor ,\label{eqn:WH2}\\
&s:=(s_1-s_2)\thinspace\textrm{mod}\thinspace Q,\label{eqn:WH3}\\
&m_p := (m_p+ s) \thinspace\textrm{mod}\thinspace p, \\
&m_q := (m_q + s)\thinspace\textrm{mod}\thinspace q, \\
&c_p := m_p^e \thinspace\textrm{mod}\thinspace p , \label{CodeExponentiationStep1}\\
&c_q := m_q^e \thinspace\textrm{mod}\thinspace q ,  \label{CodeExponentiationStep2}\\
&c :=  \left( \left( (c_p - c_q) (q^{-1}\thinspace\textrm{mod}\thinspace p)\right) \thinspace\textrm{mod}\thinspace p \right) q +c_q, \label{CodeExponentiationStep3}\\
&r := c / n . \label{CodeFloatingPointStep} 
\end{align}
\end{subequations}
Care needs to be taken in steps \eqref{eqn:WH3} and  \eqref{CodeExponentiationStep3} to avoid negative intermediate results and, 
since $n>2^{53}$, step \eqref {CodeFloatingPointStep} should include a test to avoid returning the upper limit $r=1.0$ in the IEEE double-precision floating point format.
\cite{CRTExponents}

There are 98,182,656 primes in the range $[2^{31}, 2^{32}]$.\cite{http://oeis.org/A036378} 
Since the 64-bit ciphertexts are determined by $c_p\in\mathbb{Z}_{p}$ and $c_q\in\mathbb{Z}_{q}$ using Garner's formula, we performed some statistical tests using 32-bit prime moduli $n=p$.\cite{PohligHellman1978,Beale2014}
Most primes work fine, but occasionally primes in which $p-1$ contains only small factors 
fail some of the U01 tests. 
To avoid this 
we choose $p$ and $q$ from the set of safe primes, i.e. primes $p$ for which $(p-1)/2$ is also prime. 
\cite{Schneier1994, SophieGermain} 
This does not seriously limit the scalability of the generator 
since there are 3,060,794 safe primes in the range $[2^{31} , 2^{32}]$.\cite{http://oeis.org/A211395} 
Using safe primes has the added advantage that every small odd exponent $e$ is coprime to $\phi(n)$.

Choosing $n \approx Q$ 
has the advantage that the $D$-dimensional message sequence $\{m_k, m_{k+1} .. , m_{k+D} \}$ samples $\mathbb{Z}_n^D$ nearly uniformly, 
so each message approximates a pseudorandom one-time pad-like encryption of the previous message.  
This choice reduces the total number of moduli available, but there are still over ten million independent values of $n$ that 
differ from $Q$ by less than one part in a million. The full period of each generator is then $P=(Q-1)n \approx 2^{126} \approx 8.5 \times 10^{37}$. 

One can use fast primality tests to select the primes $p$ and $q$. The Rabin-Miller 
test,\cite{millerprime, rabinprime,FergusonSchneierKohno2010,Schneier1994,Koblitz1987} which is the same as 
Algorithm P in Knuth,\cite{knuth} provides a simple probabilistic test for primality. Every odd prime $p=1+2^u t$ with $t$ odd 
satisfies one of the following conditions for every base 
$g\in \mathbb{Z}_p^*$: either $g^t \thinspace\textrm{mod}\thinspace p = 1 $, or $g^{2^j t} \thinspace\textrm{mod}\thinspace p = p-1$ for some $j$
in the range $0 \leq j < u$. A composite modulus $n$ satisfying these criteria is called a strong pseudoprime to the base $g$.  
For every odd composite, the number of bases  
for which $n$ is a strong pseudoprime  
is less than $n/4 $, so if the test is applied repeatedly with $M$ randomly chosen bases in $\mathbb{Z}_{n}^*$, the probability that a composite will pass every test is less than $4^{-M}$.\cite{Koblitz1987, knuth, millerprime, rabinprime} Better yet, the Rabin-Miller test can conclusively identify all primes below $2^{64}$ since there are no composite numbers below $2^{64}$ that are strong pseudoprimes for all of the 
twelve smallest prime bases ($g=2,3,5,7,11,13,17,19,23,29,31,37$).\cite{PomeranceSelfridgeFlagstaff1980, http://oeis.org/A014233, Zhang2001, JiangDeng2012, wolframrabinmiller}
Therefore, any number less than $2^{64}$ that passes the Rabin-Miller test for all twelve of these bases is prime. 
Likewise, any number less than $2^{32}$ that passes the Rabin-Miller test for all of the five smallest prime bases ($g = 2, 3, 5, 7, 11$) is prime. For efficiency, one first checks to see if any small primes divide the modulus before applying the Rabin-Miller test.


\section{\label{sec:Parallelization}Parallelization and Vectorization}

In a multiprocessor supercomputer environment, independent parallel pseudorandom streams can be created by assigning distinct parameter sets $\{p,q,a,e\}$ to each process. 
Message Passing Interface (MPI) calls or equivalent can be used to initialize the parallel processes with independent parameters.


Algorithm \eqref{algorithm1} can be vectorized 
to take advantage of vector processing or multiple cores available to each process.
We can use stream splitting in each independent process to return a vector of pseudorandom reals $\boldsymbol{r}$ with length $M_v$ by acting simultaneously on vectors of messages $\boldsymbol{m}$ and skips $\boldsymbol{s}$ 
which can be updated simultaneously with fixed parameters $n=p q$, $e$, $Q$, and $a$. 
The vector pseudocode 
to update the $M_v$ skips and messages, and return $M_v$ pseudorandom reals 
is 
\begin{subequations}
\begin{align}\label{vectorize}
&\boldsymbol{s} := a \boldsymbol{s} \thinspace\textrm{mod}\thinspace Q ,\\
&\boldsymbol{m} := \left(\boldsymbol{m} + \boldsymbol{s} \right)\thinspace\textrm{mod}\thinspace n ,\\
&\boldsymbol{c} :=  \boldsymbol{m}^e \thinspace\textrm{mod}\thinspace n ,\\
&\boldsymbol{r} := \boldsymbol{c}/n ,
\end{align}
\end{subequations}
although the calculation would utilize restricted primitive roots and the CRT to update the vectors using 64-bit arithmetic following algorithm \eqref{algorithm1}.
The elements of the vectors can be updated simultaneously in parallel on a vector processor, or the calculation can be shared among multiple cores available to each process using Open Multiprocessing (OpenMPI). 

To ensure that the $M_v$ sub-streams 
labelled $\gamma=0,1,..,M_v-1$
sample greatly separated subsequences, one can implement a stream splitting approach by using the jump ahead property of the skip generator 
$s_k=s_0 a^k \thinspace\textrm{mod}\thinspace Q$ 
to widely distribute the skips across the period of the skip generator.
This can be accomplished by using 128-bit arithmetic to set initial values of the elements of the skip vector to be $s^{(\gamma)}_0 = s^{(0)}_0 a^{\gamma\lfloor(Q-1)/M_v\rfloor}\thinspace\textrm{mod}\thinspace Q$ for $\gamma=0,\ldots , M_v-1$. This ensures that the message and skip sequences are all independent until the skips in successive sub-streams begin to overlap after the vector has been updated $\lfloor (Q-1)/M_v\rfloor$ times, 
 i.e. after a total of about $Q-1$ pseudorandom numbers have been generated.  Since $Q\approx 2^{63}$, this would currently take many years for a single node to accomplish. 

 

We developed and tested our code on the University of Colorado Boulder Summit supercomputer,
which uses  2.50GHz Intel Xeon E5-2680 v3 processors and 24 cores per node.\cite{RMACC} We tested the speed of the code by averaging sequences of $10^9$ pseudorandom numbers, and used  OpenMP to share the calculation across the multiple cores on each node. 
By assigning all 24 cores to each process, the code generates more than $10^8$ pseudorandom numbers per second per process for exponents as large as $e=257$. Smaller exponents $5 \leq e\leq 17$ have 24 core speeds of $1.25\times10^8$  to $1.72\times10^8$ pseudorandom numbers per second per process without affecting the quality of sequences, and eight core speeds range  from $5.6\times10^7$  to $9.3\times10^7$ pseudorandom numbers per second per process. 
By comparison, the highly optimized  
pseudorandom number generators in the Intel MKL library deliver $3$ to $6\times 10^8$ pseudorandom numbers per second per process.

\section{\label{sec:Tests}Tests}

We applied the well-established pseudorandom number test suites DIEHARD,\cite{diehard} NIST,\cite{NIST} and TestU01,\cite{U01Test} to ensure the generator passes a wide variety of tests, and calculated $\chi^2$ and the associated $p$-value for the following fourteen additional chi-squared tests.

\begin{itemize}

\item One-dimensional frequency test:\cite{knuth} We distributed sequences of pseudorandom numbers into a one-dimensional histogram with $2^{20}$ bins, and compared the histogram to a uniform Poisson distribution. 

\item Serial test in $D$=2, 3, 4, 5, and 6 dimensions:\cite{knuth} We distributed sequences of $D$ successive pseudorandom numbers $\{r_1, \ldots ,r_D\}$ into a $D$-dimensional histogram with either $2^{20}$ or $10^6$ bins, and compared the histogram to a uniform Poisson distribution. This tests for $D$-dimensional sequential correlations  in the sequence.




\item Poker test:\cite{knuth} We used groups of five pseudorandom numbers and counted the number of pairs, three-of-a-kind etc.~formed from five cards with sixteen denominations, and compared the resulting histogram to a Poisson distribution derived from the exact probabilities. This tests for a variety of five-point and shorter correlations in the sequence.

\item Collision tests:\cite{knuth} We used the pseudorandom stream to distribute $2^{14}$ balls into $2^{20}$ urns, and compared the distribution of the number of collisions with the exact distribution. We used this to test for correlations in the twenty most significant bits of each pseudorandom number, and  the most significant bit of twenty sequential pseudorandom numbers. 

\item Gaps test:\cite{knuth} We compared the histogram of the length of runs of 0's ($r\leq 0.5$) and 1's ($r>0.5$) to the exact Poisson distribution to test for correlations in the high-order and low-order  bits.

\item Max-of-t test:\cite{knuth} We compared the distribution of the maximum value among $\{r_1,r_2,\ldots ,r_t\}$ for $t=32$ with the exact probability distribution.

\item Permutations test:\cite{knuth} We compared the permutation ordering number of $t$ successive pseudorandom numbers $\{r_1,r_2,\ldots ,r_t\}$ for $t=10$, with the uniform distribution of $t!$ possibilities.

\item Fourier test:\cite{MascagniSrinivasan2000} We used a fast Fourier transform\cite{fft} to calculate the Fourier coefficients of sequences of  $M=2^{20}$ real pseudorandom numbers, 
\begin{align}
\hat{x}_k = \frac{1}{\sqrt{M}} \sum_{j=0}^{M-1} { x_j e^{2\pi i j k/M} } ,
\end{align}
where $x_j=(r_{2j} - 0.5) + i (r_{2j+1} - 0.5)$. We compared the distribution of the real and imaginary parts of  $\hat{x}_k$  with a normal distribution with zero mean and variance $1/12$. This test exposes long-range pair correlations in the pseudorandom sequence. 

\item Two-dimensional Ising model energy distribution test:\cite{beale1996, pathriabeale2011} We performed Wolff algorithm\cite{wolff} Monte Carlo simulations at the critical point of the two-dimensional Ising model on a $128\times128$ square lattice, and compared the energy histogram to a Poisson distribution derived from the exact probabilities\cite{beale1996, pathriabeale2011} Since the Wolff algorithm is based on stochastically growing fractal critical clusters that can span the system, this tests for long-range correlations in the pseudorandom sequence, and has proven to be effective at identifying weak generators.\cite{beale1996,pathriabeale2011,ferrenberglandauwong1992} 
See figure 1.

\end{itemize}
\begin{figure}[h!]
\includegraphics[width=4.5in]{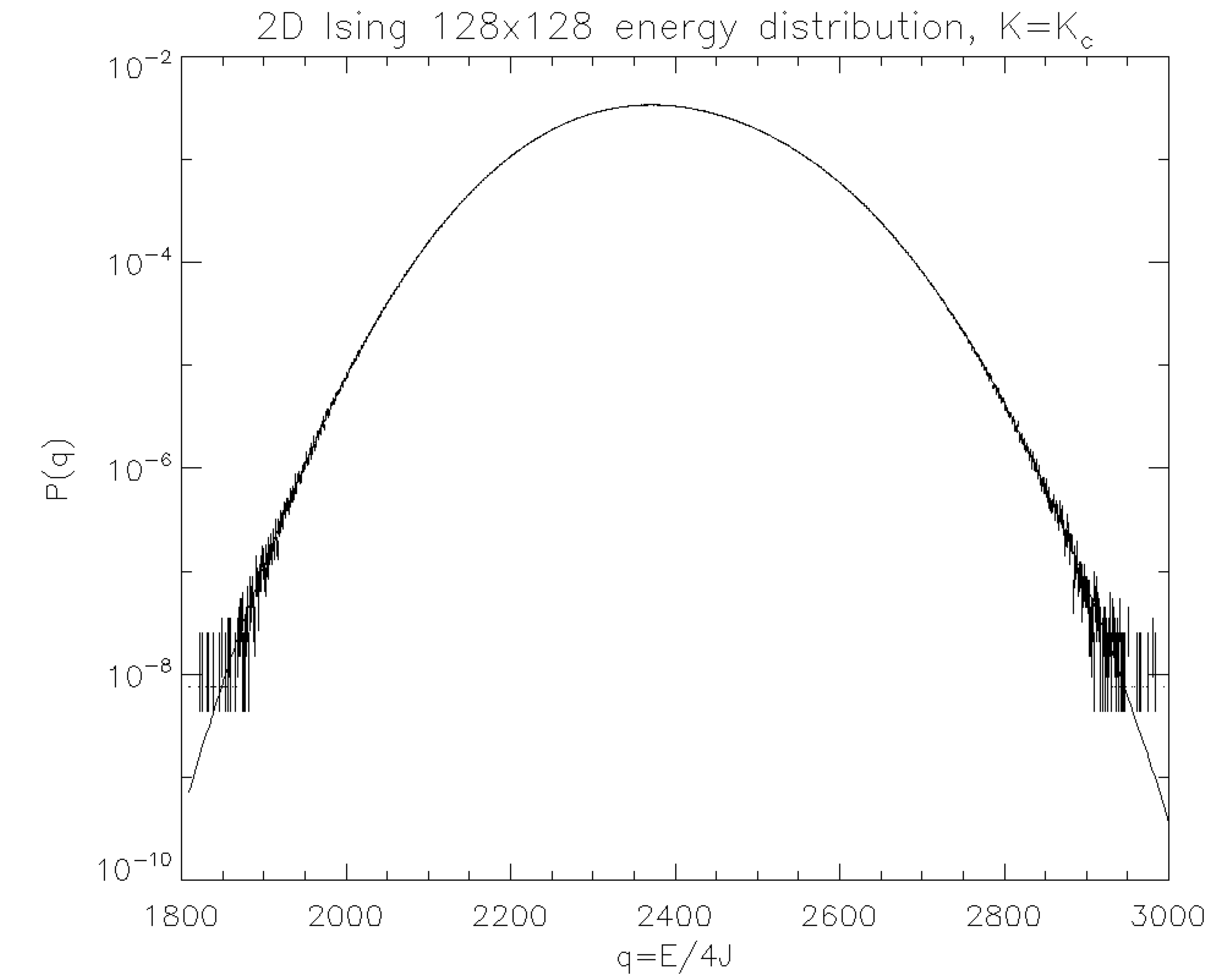}
\caption{The exact energy distribution\cite{beale1996,pathriabeale2011} for a $128 \times128$ square-lattice two-dimensional Ising model at the critical temperature (solid line) on a log scale, and the distribution calculated from $1.34\times 10^8$ configurations chosen from $4.36\times 10^8$ Monte Carlo steps per spin. The simulation was performed using  from a Wolff algorithm\cite{wolff} (error bars). The Wolff algorithm effectively eliminates critical slowing down, so the selected configurations are nearly uncorrelated with correlation time $\tau=0.44$.
The abcsissa is the energy above the ordered ground state in units of four times the coupling constant $J$. 
The simulation included 32 independent parallel processes, using approximately $10^{13}$ pseudorandom numbers generated with exponent $e=9$. 
The result was $\chi^2=990$ with 1026 degrees of freedom, for a $p$-value of $p=0.79$. 
}
\end{figure}

We tested the generator with thousands of different safe prime pairs for exponents as small as $e=3$. Every instance passed all of our correlations tests, with as many as $10^{13}$ pseudorandom numbers per test. In no case did a chi-squared test produce a $p$-value less than $10^{-6}$ or greater than $1-10^{-6}$. We also counted the number of tests that produced $p$-values less than $10^{-3}$ or greater than $1-10^{-3}$, and confirmed the number was consistent with the expected rate of one per one-thousand for each. 
We also applied the 1D frequency test, $D$-dimensional serial tests, the poker test, collisions tests, and gaps tests to the least significant bits, which also passed every test.

We confirmed that the algorithm displayed lack of correlation between streams. 
Each of $M_p$ distinct streams labeled $\alpha=0,1,.. M_p -1$ was assigned  a different prime pairs $p$ and $q$.
Our interstream correlations tests drew the pseudorandom numbers from the $M_p$ streams in the order $r_1^{(0)}, r_1^{(1)}, r_1^{(2)}, \ldots , r_1^{(N_p-1)}, r_2^{(0)}, r_2^{(1)}, \ldots$ , with various values of $M_p$. 
To test that seeding coincidences do not cause correlations, we performed the interstream correlations tests using the same primitive root $a$ in each stream, and initialized every sequence with the same values $m_0=0$ and $s_0=1$. The interstream correlations passed all of the U01 SmallCrush, Crush and BigCrush tests, even for $e=3$.


To examine the resilience of the generator, we tested various intentionally weakened versions of the generator. As noted before, the generator passes the U01 tests with a unit or constant skip for $e\geq 9$. We tested the generator with $e=1$, i.e. $c_k=m_k$ to test our use of $n\approx Q$. Since the messages nearly uniform sample of $Z_n^D$, the message sequences themselves pass gentle randomness tests such as U01 SmallCrush. To test the sensitivity of the generator to the quality of the skip sequence, we tested the skip generator with $a=3,6,7,10,11$, the five smallest primitive roots mod $Q$. The generator passes the U01 Crush tests with $e=3$ even with these intentionally bad primitive roots. 

\section{\label{sec:Conclusion}Conclusion}
We propose a class of parallel pseudorandom number generators based on a non-cryptographic RSA exponentiation cipher 
operating on 64-bit messages. The method is fully scalable based on parametrization since each process can be assigned a unique composite 
modulus $n=p q$, where $p$ and $q$ are 32-bit safe primes, and the period 
of each instance greater than $8\times 10^{37}$. 
By vectorizing the calculation, the method can produce over one-hundred million pseudorandom numbers per second on each a node of a multi-core supercomputer. We tested thousands of different pseudorandom streams, and all passed a battery of intrastream and interstream statistical tests. The C source code 
is available at 
\url{https://github.com/PDBeale/randomRSA.git}.

\section{Acknowledgements} We dedicate this paper to our late friend, mathematician Rudy L. Horne. We thank Rudy, Matt Glaser, Nick Mousouris, Ethan Neil, Robert Blackwell, John Black, David Grant  and Nick Featherstone for helpful discussions. 
This work utilized the RMACC Summit supercomputer, which is supported by the National Science Foundation (awards ACI-1532235 and ACI-1532236), the University of Colorado Boulder, and Colorado State University. The Summit supercomputer is a joint effort of the University of Colorado Boulder and Colorado State University.\cite{RMACC}


\begin{thebibliography}{5}


\bibitem{rsa1978} R. Rivest, A. Shamir, and  L. Adleman, 
Commun. ACM
\textbf{21}, 120 (1978).

\bibitem{FergusonSchneierKohno2010} N. Ferguson, B. Schneier, T. Kohno, \textit{Cryptography Engineering: Design Principles and Applications}, (Wiley, Indianapolis, 2010).

\bibitem{Schneier1994} B. Schneier, \textit{Applied Cryptography} (Wiley, New York, 1994).

\bibitem{Koshy2002} T. Koshy,  \textit{Elementary Number Theory and Applications} (Academic, San Diego, 2002).

\bibitem{Silverman2006} J.H. Silverman,  \textit{A Friendly Introduction to Number Theory} (Pearson, New York, 2006).

\bibitem{Koblitz1987} N. Koblitz, \textit{A Course in Number Theory and Cryptography} (Springer-Verlag, New York, 1987).



\bibitem{NumericalRecipes1992} W.H. Press, S.A. Teukolsky, W.T. Vetterling, and B.P. Flannery, \textit{Numerical Recipes: The Art of Scientific Computing}, (Cambridge, New York, 1992), 2nd ed.  

\bibitem{Rijmen} V. Rijmen, A. Bosselaers, and P. Barreto. Optimized ANSI C code for the Rijndael cipher (now AES), 2000. Public domain software.


\bibitem{Random123} J.K. Salmon, M.A. Moraes, R.O. Dror, and D.E. Shaw, 
Proceedings of the International Conference for High Performance Computing, Networking, Storage and Analysis (SC11), New York, NY: ACM, 2011. doi:10.1145/2063384.2063405 .

\bibitem{U01Test} P. L'Ecuyer and R. Simard, 
ACM Trans. Math. Software 
\textbf{33}, 22 (2007); see \url{http://www.iro.umontreal.ca/~simardr/testu01/tu01.html}.






\bibitem{Shparlinski2000} I.E. Shparlinski, Math. Comp. \textbf{70}, 801 (2000).


\bibitem{Coddington} P.D. Coddington, 
Northeast Parallel Architecture Center. Paper 13. \url{https://surface.syr.edu/npac/13/} (1997)


\bibitem{BaukeMertens2007}H. Bauke and S. Mertens, Phys. Rev. E \textbf{75}, 066701 (2007).



\bibitem{MascagniSrinivasan2000}
M. Mascagni and A. Srinivasan, 
ACM Trans. Math. Software
\textbf{26}, 436 (2000).

\bibitem{sprng} The Scalable Parallel Random Number Generators Library (SPRNG), \url{http://www.sprng.org}.



\bibitem{mascagni1998} M. Mascagni, 
Parallel Comput. 
\textbf{24}, 923 (1998).  


\bibitem{mascagni1995a} M. Mascagni, M. L. Robinson, D. V. Pryor and S. A. Cuccaro, 
{Lec. Notes Statistics} \textbf{106}, 263 (1995).

\bibitem{mascagni1995b} M. Mascagni, S. A. Cuccaro, D. V. Pryor and M. L. Robinson,  
{J. Comp. Phys.} \textbf{119}, 211 (1995).

\bibitem{MersenneTwister} M. Matsumoto and T. Nishimura, 
ACM Trans. Model. Comput. Sim. \textbf{8(1)}, 3 (1998).

\bibitem{knuth}D. Knuth, \textit{The Art of Computer Programming}, vol. 2 (Addison-Wesley, Reading, Massachusetts, 1999).

\bibitem{ZierlerBrillhart}N. Zierler and J. Brillhart, Info. and Control \textbf{13}, 541 (1968).

\bibitem{Merkle1978} R.C. Merkle,  Communications of the ACM. \textbf{21 (4)}: 294 (1978).

\bibitem{DiffieHellman1976} W. Diffie, and M.E. Hellman, IEEE Transactions on Information Theory. \textbf{22 (6)}, 644  
(1976).

\bibitem{SinghCodeBook}S. Singh, \textit{The Code Book}, (Doubleday, New York, 1999).

\bibitem{Shor} P.W. Shor, SIAM J. Comp. \textbf{26}, 1484 (1977); \url{https://arxiv.org/abs/quant-ph/9508027}.









\bibitem{WichmanHill1982} B.A Wichmann, and I.D. Hill, I.D,  Appl. Stat. \textbf{31}, 188 (1982).

\bibitem{BratleyFoxSchrage1987} P. Bratley, B. L. Fox,, and L.E. Schrage,. \textit{A Guide to Simulation}, 2nd ed..  (Springer-Verlag, New York, 1987).

\bibitem{lecuyer1988} P. L'Ecuyer, Commun. ACM \textbf{31}, 741 (1988).



\bibitem{LecuyerBlouinCouture1993} P. L'Ecuyer. F.-O. Blouin, and R. Couture, ACM Trans. Model. Comput. Simul.. \textbf{3}, 87 (1993). They recommend primitive root $a=2307085864$.

\bibitem{SezginSezgin2013} F. Sezgin and T. M. Sezgin, Comp. Phys. Comm. \textbf{184}, 1889  (2013). Not all of the multipliers in Table 4 are primitive roots mod $Q=2^{63}-25$. We have confirmed that multipliers \hfil\break
$a=\{3157107955,3163786287, 3200261722, 3211103532,3338736601, 3423977237, 3465965455, 3474009732, 3512424704\}$ are primitive roots mod $Q$ and pass all U01 Crush and BigCrush tests.








\bibitem{SophieGermain}This makes $(p-1)/2$ a Sophie Germain prime.


\bibitem{PohligHellman1978} S. Pohlig and M. Hellman, 
IEEE Transactions on Information Theory IEEE Trans. Inform. Theory \textbf{(24)}, 106 (1978).


\bibitem{Beale2014} P.D. Beale, \url{https://arxiv.org/abs/1411.2484}.


\bibitem{CRTExponents} One might also use the CRT and Fermat's Little Theorem to implement a large exponent $e$ without sacrificing execution speed as long as $e$ has small remainders modulo $p-1$ and $q-1$.
If $e_p$ is a small chosen exponent coprime to $p-1$, and $e_q$ is a \textit{different} small chosen exponent coprime to $q-1$, then using $c_p=m_p^{e_p} \thinspace \textrm{mod} \thinspace p$ and 
$c_q=m_q^{e_q} \thinspace \textrm{mod} \thinspace q$ in algorithm steps  \eqref{CodeExponentiationStep1} and \eqref{CodeExponentiationStep2} gives a ciphertext $c=m^e \thinspace \textrm{mod} \thinspace n$ with a large exponent $e$ that obeys $e_p=e \thinspace \textrm{mod} \thinspace (p-1)$ and  $e_q=e \thinspace \textrm{mod} \thinspace (q-1)$. If needed, the value of $e$ can be determined using the CRT.
 
 
\bibitem{http://oeis.org/A036378} \textit{The On-Line Encyclopedia of Integer Sequences}, \url{http://oeis.org/A036378}.

\bibitem{http://oeis.org/A211395} \textsl{The On-Line Encyclopedia of Integer Sequences}, \url{http://oeis.org/A211395}; \url{http://oeis.org/A211397}.

\bibitem{millerprime} G.L. Miller,  
J. Comp. Sys. Sci.
\textbf{13}, 300 
(1976).

\bibitem{rabinprime} M.O. Rabin, 
J. Number Theory
\textbf{12}, 128
 (1980).



\bibitem{PomeranceSelfridgeFlagstaff1980} C. Pomerance, J. L. Selfridge and S. S. Wagstaff, Jr., 
Math. Comp
\textbf{35}, 1003 (1980).

\bibitem{http://oeis.org/A014233} \textsl{The On-Line Encyclopedia of Integer Sequences} \url{http://oeis.org/A014233}.



\bibitem{Zhang2001} Zhenxiang Zhang, 
Math. Comp. \textbf{70}, 863 (2001).

\bibitem{JiangDeng2012}
Y. Jiang and Y. Deng, \url{http://arxiv.org/abs/1207.0063v1}.

\bibitem{wolframrabinmiller}See \url{http://mathworld.wolfram.com/Rabin-MillerStrongPseudoprimeTest.html}.








\bibitem{diehard} G. Marsaglia, DIEHARD: a battery of tests of randomness (1996); 
see \url{http://stat.fsu.edu/pub/diehard/}.

\bibitem{NIST} A. Rukhin, J. Soto, J. Nechvatal, M. Smid, E. Barker, S. Leigh, M. Levenson, M. Vangel, D. Banks, A. Heckert, J. Dray, and S. Vo, 
NIST special publication 800-22, National Institute of Standards and Technology (NIST), Gaithersburg, Maryland, USA, 2001; see \url{http://csrc.nist.gov/rng/}.

\bibitem{fft} J.W. Cooley, J.W. Tukey, 
Math. Comp. \textbf{19}, 297 (1965). .


\bibitem{beale1996} P.D. Beale, {Phys. Rev. Lett.} \textbf{76}, 78 (1996).


\bibitem{pathriabeale2011} R.K. Pathria and P.D. Beale, \textit{Statistical Mechanics} 3rd ed., (Academic, Boston, 2011).



\bibitem{wolff} U. Wolff, {Phys. Rev. Lett.} \textbf{62}, 361 (1989).

\bibitem{ferrenberglandauwong1992} A.M. Ferrenberg, D.P. Landau and Y.J. Wong, {Phys. Rev. Lett.} \textbf{69}, 3382 (1992).








\bibitem{RMACC} Jonathon Anderson, Patrick J. Burns, Daniel Milroy, Peter Ruprecht, Thomas Hauser, and Howard Jay Siegel. 2017. Deploying RMACC Summit: An HPC Resource for the Rocky Mountain Region. In Proceedings of PEARC17, New Orleans, LA, USA, July 09-13, 2017, 7 pages. DOI: 10.1145/3093338.3093379



\end{thebibliography}
\end{document}